\documentclass[lettersize,journal]{IEEEtran}
\usepackage{amsmath,amsfonts}
\usepackage{algorithmic}
\usepackage{algorithm}
\usepackage{array}
\usepackage[caption=false,font=normalsize,labelfont=sf,textfont=sf]{subfig}
\usepackage{textcomp}
\usepackage{stfloats}
\usepackage{url}
\usepackage{verbatim}
\usepackage{graphicx}
\usepackage{cite}
\newtheorem{theorem}{Theorem}

\newtheorem{lemma}{Lemma}
\newtheorem{remark}{Remark}

\usepackage{mathtools}
\usepackage{multicol}
\usepackage{xcolor}
\usepackage{multirow}
\usepackage{epstopdf}

\begin{document}

\title{Beam Scheduling in Millimeter Wave Networks Using the Whittle Index}

\author{Mandar R. Nalavade, Ravindra S. Tomar, Gaurav S. Kasbekar,~\IEEEmembership{Member,~IEEE,} 
\thanks{The authors are with the Department of Electrical Engineering, Indian Institute of Technology (IIT) Bombay, Mumbai, 400076, India. Their email addresses are 22d0531@iitb.ac.in, 30005388@iitb.ac.in, and gskasbekar@ee.iitb.ac.in. The work of all three authors has been supported in part by the project with code RD/0121-MEITY01-001.}}



\maketitle

\begin{abstract}
We address the problem of beam scheduling for downlink transmissions in a single-cell millimeter wave (mmWave) network. The cell contains a mmWave base station (mBS) and its associated users. 
At the end of each time slot, a packet arrives into the queue of a user at the mBS with a certain probability. A holding cost is incurred for the packets stored in a user's queue at the mBS in every time slot. The number of simultaneous beams that the mBS can form to different users is less than the number of associated users. Also, a cost is incurred whenever a beam is formed from the mBS to a user. In a given time slot, a packet transmitted from the mBS to a user that has been assigned a beam is successfully received (respectively, not received) if the channel quality between the mBS and the user is good (respectively, bad). In every time slot, the mBS needs to assign the available beams to a subset of the users, in order to minimize the long-run expected average cost. This problem can be modeled as a restless multi-armed bandit problem, which is provably hard to solve. We prove the Whittle indexability of the above beam scheduling problem and propose a strategy to compute the Whittle index of each user. In each time slot, our proposed beam scheduling policy assigns beams to the users with the smallest Whittle indices. Using extensive simulations, we show that our proposed Whittle index-based beam scheduling policy significantly outperforms several  scheduling policies proposed in prior work in terms of the average cost, average delay, as well as energy efficiency.
\end{abstract}


\section{Introduction} \label{section_1_intro}
\IEEEPARstart{T}{he} emergence of technologies like artificial intelligence (AI), virtual reality (VR), augmented reality (AR), Internet of Things (IoT), and big data analytics, has led to the deployment of a plethora of new bandwidth-intensive applications, significantly increasing the data traffic in wireless networks \cite{xiao2017millimeter}. Future networks are expected to handle unprecedented traffic volumes, and this exponential increase in data traffic demand has  made it challenging for service providers to address the resulting bandwidth shortage \cite{xiao2017millimeter}, \cite{rappaport2013millimeter}. While sub-6 GHz bands have limited available bandwidth, the millimeter wave (mmWave) spectrum provides abundant bandwidth and is capable of supporting data rates of several gigabits per second per user. MmWave frequencies exhibit distinct propagation characteristics compared to sub-6 GHz frequencies, e.g., mmWave communications are more susceptible to blockages,  rapid channel variations, high propagation losses, and atmospheric absorption \cite{bai2014coverage}. A notable advantageous aspect of the shorter wavelengths of mmWaves is that they allow for a denser packing of antenna elements within the same physical dimensions compared to sub-6 GHz frequencies. These large antenna arrays can leverage \emph{beamforming} to compensate for path loss and establish communication links with adequate signal-to-noise ratios (SNR) \cite{el2014spatially}. The beams generated by arrays of antenna elements  in mmWave systems are narrow, leading to focused signal transmission and creating the potential to spatially multiplex several beams, enabling simultaneous service to multiple users from a single mmWave base station (mBS).

In a dense mmWave cell, the number of users associated with a mBS often exceeds the number of simultaneous beams the mBS can form, as the number of beams is constrained by the number of available radio frequency (RF) chains at the mBS \cite{dartmann2013application}. Since the mBS cannot serve all its users simultaneously with the limited number of available beams, \emph{beam scheduling}, i.e., selecting a subset of users to which beams will be assigned in a given time slot,  is essential.  At the mBS, packets arrive from the core network for transmission over the wireless channel and are queued in buffers or packet queues-- typically, there are separate queues for the packets of different users. Given the packet queues at the mBS for different users, the beam scheduler at the mBS determines as to which users should be allocated beams in each time slot. 

Several studies have addressed the problem of beam scheduling in mmWave networks. For example, beam scheduling in multi-cell scenarios was studied in \cite{dartmann2013application, sha2019least, sha2020graph}, and beam-aware scheduling with the objective of saving the user equipment (UE) power in mmWave networks was studied in \cite{weng2018beam} and \cite{pan2023beam}. The beam selection problem was studied in \cite{wang2021multi, zou2021novel, jiang2018joint, liu2018decentralized, hegde2019matching, ahn2022machine, yang2021multi, yang2020mmmuxing, wei2017pose, vuckovic2023paramount}. A many-to-many beam alignment protocol was presented in \cite{jog2019many}. The work \cite{gopalam2024joint} investigated joint beam allocation and scheduling, whereas \cite{xu2019ping} investigated the user selection and beam allocation problem.  The problem of joint user and beam selection was analyzed in \cite{cheng2020low}; also, \cite{boljanovic2024joint}, \cite{singh2024joint} studied joint user association and beam scheduling in mmWave networks. A method to jointly optimize user scheduling and beam scheduling was proposed in \cite{sha2023versatile}. In \cite{ju2023deep}, a joint beam allocation and relay selection scheme was proposed to mitigate blockages and enhance the total transmission rate in mmWave vehicular networks. A machine learning (ML) based beam allocation algorithm for mmWave networks was proposed in \cite{kose2024multi}. In \cite{ju2024energy}, the Dueling Double Deep Recurrent Q-Network (D3RQN) algorithm was proposed for beam allocation. Also, various packet scheduling techniques, including Longest-Queue-First (LQF) \cite{joo2009understanding}, Max-Weight Scheduling (MWS) \cite{kadota2018scheduling, tsanikidis2022randomized, wang2022scheduling}, Weighted Fair Queuing (WFQ) \cite{parekh1993generalized, rashed2010comparative, pan2022deep, li2016analytical}, and Random scheduling \cite{kadota2019minimizing, tsanikidis2022randomized}, can potentially be used to address the beam scheduling problem. However, to the best of our knowledge, the powerful \emph{Whittle index} technique \cite{whittle1988restless} has not been used for beam scheduling in a wireless network in prior work. This is the space in which we contribute in this paper.

The Whittle index, originally proposed  in \cite{whittle1988restless}, is a popular technique for decision-making in stochastic systems, and has been applied to effectively solve problems  across a wide variety of domains, including minimizing the age of information (AoI) \cite{kadota2019scheduling, hsu2019scheduling, maatouk2020optimality, tripathi2024whittle}, scheduling in wireless networks \cite{borkar2017opportunistic, wei2021two, karthik2022scheduling}, allocating jobs
to processors in an egalitarian processor sharing setup \cite{borkar2022whittle}, user association in mmWave networks \cite{singh2022user, nalavade2024whittle, tomar2024user}, pilot allocation in massive multiple-input multiple-output (MIMO) systems \cite{yang2020restless}, beam resource allocation in radar systems for multitarget tracking \cite{hao2024neural}, etc.

This paper considers downlink (DL) transmission in a single-cell mmWave network, wherein a mBS serves its associated users. Time is divided into equal slots, and at the end of every time slot, a packet arrives into the queue of a user at the mBS with a certain probability. A holding cost is incurred for the packets stored in a user's queue at the mBS in every time slot. The number of simultaneous beams that the mBS can form is less than the number of users in the cell. Also, a cost is incurred whenever a beam is formed from the mBS to a user. In each time slot, the mBS needs to assign the available beams to a subset of the users in such a way that each beam is assigned to at most one user. The channel quality between the mBS and a user in a given time slot is a random variable. Also, a packet transmitted from the mBS to a user that has been assigned a beam is successfully received (respectively, not received) if the channel quality between the mBS and the user is good (respectively, bad). Our objective is to design a beam scheduling policy, which in every time slot selects a subset of users to which beams are assigned, so as to minimize the expected long-run time-averaged total holding cost of the packets of all the users; note that such a policy would lead to low average delays experienced by packets in the system. This problem is an instance of the  \emph{restless multi-armed bandit}  (RMAB) problem, which is widely used to model scenarios involving sequential decision-making under given constraints.  RMAB problems are provably hard to solve \cite{papadimitriou1994complexity}. Using an idea by Whittle \cite{whittle1988restless}, we relax the hard per-stage constraint that the number of users to which beams are assigned in a time slot must equal the number of available beams, to a long-run time-averaged constraint. Subsequently, we use a Lagrangian multiplier technique to transform the problem into an unconstrained problem and then decouple the problem into separate Markov Decision Processes (MDPs) for different users.
We prove the Whittle indexability \cite{whittle1988restless} of the beam scheduling problem and propose a strategy to compute the Whittle index of each user in the network. In each time slot, our proposed beam scheduling policy assigns beams to the users with the smallest Whittle indices. Using extensive simulations, we show that our proposed Whittle index-based beam scheduling policy significantly outperforms several scheduling policies proposed in prior work in terms of the average cost, average delay, as well as energy efficiency. 

Note that, in general, it is challenging to prove the Whittle indexability of a RMAB problem. This paper's main contribution is a rigorous proof of the fact that the beam scheduling problem in a mmWave network is Whittle indexable. Moreover, to the best of our knowledge, this paper is the first to address the beam scheduling problem in wireless networks using the Whittle index technique.

The remainder of this paper is structured as follows. Section \ref{related_work} presents a review of related prior work. Section \ref{system_model} describes the system model and problem formulation. The dynamic programming equation for the problem is introduced in Section \ref{value_Function_sec}. In Section \ref{Structural_prop}, we establish some useful structural properties of the value function. Section \ref{Section7_thres} proves that the optimal policy is a threshold policy. In Sections \ref{Section8_wi} and \ref{Section9_comp}, we prove the Whittle indexability of the problem and propose a method for computing Whittle indices, respectively. Section \ref{other_policies} briefly describes some scheduling algorithms proposed in prior work, and Section \ref{simulations} provides a comparison, via simulations, of the performance of our proposed policy with that of the policies described in Section \ref{other_policies}. Finally, Section \ref{conclusion} provides the conclusion of the paper and some future research directions.

\section{Related Work} \label{related_work}
We first review  prior work on beam scheduling in mmWave and massive MIMO networks. In  \cite{dartmann2013application}, beam scheduling algorithms for multi-cell networks were presented, aiming to enhance the system performance by maximizing the total throughput or improving the minimum Signal-to-Interference-Plus-Noise Ratio (SINR) among all the users. A graph-theory-based framework was employed to solve the combinatorial optimization problem corresponding to beam scheduling. In \cite{sha2019least}, the beam scheduling problem in the time-domain for mmWave systems with two cells was investigated. The system model allows each cell to transmit multiple beams concurrently, with the assumption that beams are paired one-to-one between neighboring cells. The objective is to minimize pairwise collision time slots while satisfying the quality of service (QoS) requirements for each beam. The authors formulated this as a combinatorial optimization problem and proposed a recursive algorithm, the Least Pairwise Collision (LPC) algorithm, to solve it. In \cite{sha2020graph}, to mitigate strong inter-cell interference (ICI) in mmWave networks, a time-domain beam scheduling problem was modeled using graph theory, with cells represented as nodes. The study introduced a Least Beam Collision (LBC) algorithm that minimizes beam collisions. This is achieved by ensuring that adjacent cell beams do not point to local users simultaneously. The LBC algorithm was shown to achieve the global minimum beam collision solution, enhancing network reliability and the sum rate through reduced strong ICI. In \cite{weng2018beam}, a beam-aware dormancy technique was proposed for UE power conservation in mmWave networks. It allocates consecutive time slots to a UE upon the reception of a reference signal and allows the UE to enter a dormant mode until the next signal, conserving power when no data is transmitted. The technique was analyzed using a Markov chain model and  an integer linear programming problem with a power-saving constraint was formulated. In \cite{wang2021multi}, the authors modeled the beam selection problem in mmWave vehicular networks as a multi-agent multi-armed bandit problem. A multi-agent reinforcement learning beam selection algorithm was proposed, with each base station (BS) acting as an agent. Each BS learns the Q-values of actions in its action space while considering the actions of the other BSs, leading to a reduced probability of overlapping beams. Additionally, a modified combinatorial upper confidence bound (CUCB) algorithm was proposed, which shows better performance in respect of both throughput as well as the probability of overlapping beam selection, compared to benchmark policies. In \cite{zou2021novel}, the problem of joint user scheduling and beam selection so as to maximize the sum rate was formulated for DL mmWave networks. To address this non-convex problem, two algorithms were introduced: a Whale Optimization Algorithm (WOA)-based scheme and a ML-based scheme. A Lyapunov-drift-based optimization framework was developed in \cite{jiang2018joint} to solve the joint user scheduling and beam selection problem in downlink massive MIMO networks. To guarantee queue stability, a MWS policy was employed. The formulated MWS joint scheduling problem, initially a mixed-integer programming (MIP) problem, was relaxed to a multi-convex problem and then solved using a block coordinate update (BCU) algorithm.

The problem of decentralized beam pair selection for multi-beam concurrent transmission in mmWave networks was studied in \cite{liu2018decentralized}, and a Heterogeneous Multi-Beam Cloud Radio Access Network (HMBCRAN) architecture was proposed to ensure coverage and mobility. Based on HMBCRAN and candidate beam pair links, a beam pair selection optimization problem was formulated to maximize the network sum rate. The work in \cite{hegde2019matching} studied beam selection in high-dimensional mmWave multi-user systems, aiming to maximize the sum rate. The authors modeled this as a user-beam matching problem and proposed two algorithms: deferred acceptance and matching with externalities. While deferred acceptance is a low-complexity algorithm, it neglects multi-user interference, leading to a sub-optimal sum rate. Matching with externalities addresses this by considering interference effects. In \cite{ahn2022machine}, a machine learning-based vision-aided beam selection (ML-VBS) scheme was proposed for mmWave indoor systems with the objective of reducing the beam selection overhead. Given the users' angle estimates and the limited number of RF chains at the BS, the proposed scheme employs two sequential deep neural networks for joint user and beam selection under minimum rate constraints. To enhance spatial reuse in 60 GHz mmWave networks, the Multi-Dimensional Spatial Reuse (MDSR) algorithm was proposed in \cite{yang2021multi, yang2020mmmuxing}. The proposed MDSR algorithm  considers beam imperfections and reflections. By leveraging a conflict graph, MDSR optimizes the access point (AP) association, beam selection, and user scheduling to improve the spatial reuse. This approach identifies the optimal AP-user-beam combinations to minimize the interference in each scheduling cycle. To provide coverage and mobility support at very high bit rates, \cite{wei2017pose} proposed a 60 GHz network architecture known as pose information assisted (Pia) network. Pia's operation comprises two stages: sensing and execution. In the initial stage, Pia uses compressive sensing techniques to estimate the positions of the APs and reflectors. In the execution stage, it selects the APs and beams. The AP/ beam selection algorithms run whenever new pose data is received, and Pia notifies the AP of a new beam selection only if the algorithm's output differs from the previous result. In \cite{vuckovic2023paramount}, PARAMOUNT, a generalized deep learning method, was proposed for beam selection in mmWave networks using sub-6 GHz band channel state information (CSI). A non-convex optimization problem formulation was adopted for beam selection. To achieve generalization, the neural network extracted path cluster information from sub-6 GHz band CSI and mapped it to the mmWave band. A novel augmented discrete angular-delay profile (ADADP) technique was introduced to generate a high-resolution semantic representation of path clusters in the angular-delay domain. Leveraging the visual nature of ADADP, a convolutional neural network (CNN) was employed for beam selection. A many-to-many beam alignment protocol known as BounceNet was proposed in \cite{jog2019many}. BounceNet maximizes the spatial reuse by aligning AP and client beams, enabling multiple concurrent mmWave links in a narrow space. This dense spatial reuse scales the mmWave network throughput with the number of clients. A beam-aware scheduling technique-- discontinuous reception beam-aware scheduling (D-BAS)-- was introduced in \cite{pan2023beam} for mmWave networks with discontinuous reception and dual-connected UEs. D-BAS operates in two phases: beam decision, where two beams covering the most DL data requirements are selected, and beam exchange, where the adjacent BS beam scheduling is adjusted. This two-step process allows UEs to reduce the wake time and benefit from dual connectivity. Optimal joint beam allocation and user scheduling algorithms for multi-user mmWave networks were presented in \cite{gopalam2024joint}. In particular, sand-filling algorithms with linear complexity and proven optimality were proposed. 
  
In \cite{xu2019ping}, a joint optimization framework for beam selection and allocation in mmWave networks was presented, which considers beam collision as well as inter-user interference. The scenario was formulated as a non-convex combinatorial optimization problem. To solve it, a ping-pong-like algorithm based on hybrid particle swarm optimization and simulated annealing was proposed. A framework for the joint user and beam selection problem in a multi-user discrete lens array mmWave network was proposed in \cite{cheng2020low}. To optimize the beam and user selection jointly, a stable-matching-based scheme and a greedy scheme were proposed. In \cite{boljanovic2024joint}, a multi-step optimization algorithm for jointly performing user association and beam scheduling in a multi-beam interference system was proposed. The problem was formulated as a linear optimization problem, and the proposed algorithm was shown to outperform baseline user association schemes. The work in \cite{singh2024joint} investigated the problem of jointly selecting the UE and beam for each AP for concurrent transmissions in a mmWave network containing a number of APs. The objective was to maximize the sum of weighted rates of the UEs. The problem was shown to be NP-complete, and a Markov Chain Monte Carlo (MCMC)-based and a local interaction game (LIG)-based UE and beam selection algorithm were proposed and shown to achieve the optimal solution asymptotically for the problem. Also, two fast heuristic algorithms, Novel Greedy UE and Beam Selection (NGUB) 1 and 2, were proposed for solving the problem. In \cite{sha2023versatile},  a near interference-free  (NIF) scheduling scheme for jointly optimizing user scheduling and beam scheduling was presented. The proposed scheme accommodates multiple optimization objectives, such as maximizing the sum rate and ensuring the minimum user rate. The authors considered user scheduling across time, frequency, and spatial domains under hybrid analog-digital beamforming constraints. Leveraging the directional beamforming feature, they derived a solution to the resource allocation optimization problem. In \cite{ju2023deep}, the joint beam allocation and relay selection (JoBARS) problem in mmWave vehicular networks was addressed, and a deep reinforcement learning (DRL)-based method was proposed to solve it. The proposed DRL-based JoBARS method enables the mBS to rapidly determine the optimal beam allocation and relay selection, which maximizes the total vehicular transmission rate, based on real-time traffic patterns. In \cite{kose2024multi}, an ML-based algorithm called Multi-Agent Context Learning (MACOL) was proposed for beam allocation. The proposed algorithm extends the multi-armed bandit framework by incorporating context learning. This allows the ML agents to learn potential transmission interference based on shared contexts among distributed agents. The proposed approach demonstrated that utilizing knowledge of the status of neighboring beams enables the ML agent to identify and mitigate potential interference with ongoing transmissions. In \cite{ju2024energy}, the D3RQN algorithm was proposed for the joint optimization of beam allocation, relay vehicle selection, jammer vehicle selection, and transmission power selection.
However, none of the above works \cite{dartmann2013application, sha2019least, sha2020graph, weng2018beam, wang2021multi, liu2018decentralized, ahn2022machine, zou2021novel, hegde2019matching, yang2021multi, yang2020mmmuxing, wei2017pose, jog2019many, cheng2020low, vuckovic2023paramount, xu2019ping, jiang2018joint, pan2023beam, gopalam2024joint, boljanovic2024joint, singh2024joint, sha2023versatile} addresses the beam scheduling problem in wireless networks using Whittle index theory.

We now review literature on various packet scheduling algorithms, such as LQF, MWS, WFQ, and the random scheduling policy, which determine as to which queue(s) from a set of queues to schedule for transmission in each time slot. These algorithms can be used for beam scheduling in wireless networks. In \cite{ joo2009understanding}, a wireless network was represented as a graph, and the capacity region of the widely studied sub-optimal scheduling policy, Greedy Maximal Scheduling (GMS), also referred to as LQF, was analyzed. GMS selects wireless links in decreasing order of backlog, subject to interference constraints. In \cite{kadota2018scheduling}, various scheduling policies were explored to minimize the AoI in broadcast wireless networks, including the MWS policy, randomized policy, and Whittle-based scheduling policy. In \cite{tsanikidis2022randomized}, the real-time traffic scheduling problem in wireless networks, using efficiency ratio as the performance metric, was investigated under a conflict graph interference model and fading channels. An unknown finite-state Markov chain was used to model both the traffic process and the fading process. To address the traffic scheduling problem, MWS was employed, resulting in an efficiency ratio of $\frac{1}{2}$. In contrast, a randomized scheduling algorithm achieved an efficiency ratio greater than $\frac{1}{2}$. In \cite{wang2022scheduling}, scheduling techniques were explored to minimize the long-term average AoI in a multi-user single-destination IoT system, subject to a constraint on the number of retransmissions. A low-complexity slot-based max-weight (SBMW) policy was proposed for minimizing the long-run average AoI, and an infinite-horizon MDP was decomposed into finite-state MDPs for each frame in the case of retransmission with feedback. In \cite{rashed2010comparative}, WFQ, along with first-in-first-out (FIFO) and priority queuing, were analyzed and compared for three services: Voice over Internet Protocol (VoIP), videoconferencing, and file transfer. The analysis considered key performance metrics such as dropped traffic, received traffic, and end-to-end packet delivery delay. Simulations were conducted for each scheduling algorithm to evaluate its performance in the following scenarios: traffic drop, file reception, voice data reception, and video conferencing.  WFQ was shown to outperform the other two algorithms in all of these scenarios. In \cite{pan2022deep}, a WFQ-based deep reinforcement learning algorithm was proposed for G/G/1/K parallel queuing systems.  In \cite{li2016analytical}, a Markov model was used to evaluate the availability of idle radio resources in Long-Term Evolution (LTE) for vehicular safety services and to assess the impact of these safety applications on conventional LTE users. It was proposed that the idle radio resources in LTE be reserved for vehicular safety services. On these reserved LTE resources, the WFQ algorithm was used to schedule beacons for safety services. In Section \ref{simulations}, we show via simulations that our proposed Whittle index-based policy for the beam scheduling problem in mmWave networks outperforms the LQF, MWS, WFQ, and random scheduling policies reviewed above, in terms of the average cost, average delay, as well as energy efficiency.

The work in \cite{hao2024neural} investigated beam allocation for multitarget tracking in MIMO radar systems. The problem was formulated as a RMAB problem and the authors proposed a neural network-based Whittle index (NNWI) policy to solve it. This policy leverages  DRL to approximate the Whittle index, thereby achieving the optimization of the Bayesian Cramér-Rao lower bound (BCRLB) for multiple targets. Simulation results showed that the NNWI policy outperforms the myopic, REINFORCE, and amortized Q-learning policies. In contrast to \cite{hao2024neural}, which applied the Whittle-based policy for beam resource allocation in the context of multitarget tracking in MIMO radar systems, this paper proposes a Whittle index-based policy for beam scheduling in the context of DL transmissions in a single-cell mmWave network. Consequently, the system model and results presented in this paper are entirely different from those in \cite{hao2024neural}.

In \cite{buyukkoc1985cmu}, the problem of allocating a single server among $N$ queues with an arbitrary arrival distribution of new customers and a geometric service time distribution was considered. The $c\mu$ rule was used to solve the problem, where the customer holding cost associated with each queue $i \in \{1,\ldots,N\}$ is $c_i$, and $\mu_i$ is the serving rate of queue $i$. The $c \mu$ rule denotes the policy that assigns the server to a customer in queue $i$ at time $t$ if $c_i \mu_i = \max \left \{c_j \mu_j |X_j(t) >0 \right\}$, where $X_j(t)$ is the length of the $j^{th}$ queue at time $t$. There are some similarities between the system model proposed in this paper and that in \cite{buyukkoc1985cmu}; however, the former is a significant generalization of the latter.  For example, the packet holding cost considered in our paper is an arbitrary convex non-decreasing function of the number of packets, whereas the holding cost is assumed to be linear in \cite{buyukkoc1985cmu}. Additionally, in our paper, a cost is incurred whenever a beam is formed from the mBS to a user, and this cost can be different for different users. In contrast, no such cost is incurred in the system model in \cite{buyukkoc1985cmu}. Due to the above generalizations, the $c\mu$ rule proposed in \cite{buyukkoc1985cmu} is not guaranteed to be optimal under the system model in this paper. Hence, we address the problem formulated in this paper using the theory of Whittle index.  We have shown that the problem is Whittle indexable, and to the best of our knowledge, this paper is the first to devise a Whittle index-based policy for the beam scheduling problem.

\section{System Model And Problem Formulation} \label{system_model}
\begin{figure}
\centerline{\includegraphics[width=\linewidth]{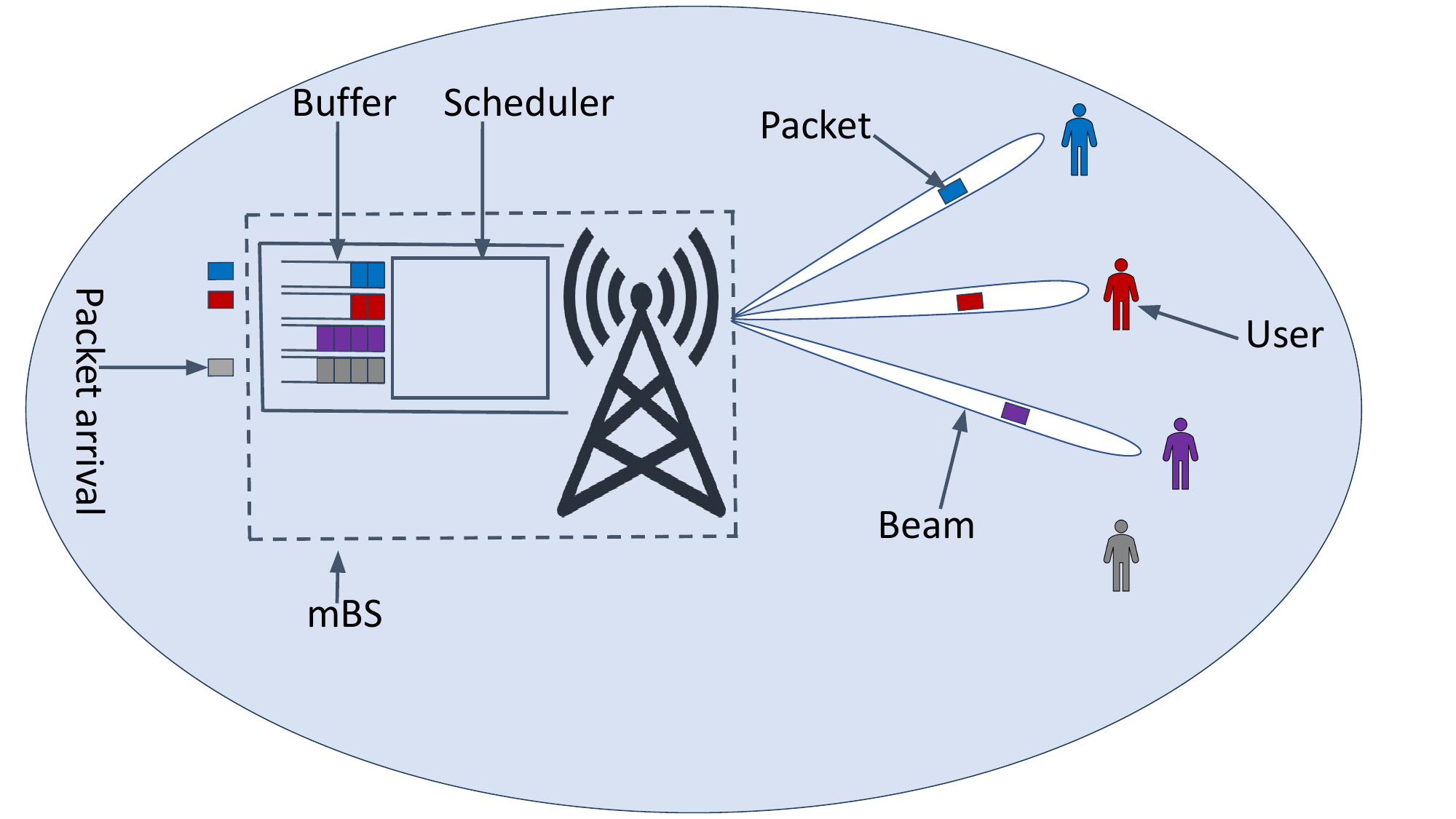}}
\caption{The figure shows an example network with $K=4$ and $B=3$.}
\label{fig:sys_model}
\end{figure}

Consider a mmWave network in which there is a single cell containing a mBS and $K \geq 2$ associated users. Time is divided into slots of equal durations, say $n \in \{0, 1, 2, \ldots \}$. The mBS has $B$ RF chains, each of which can form one beam, where $1\leq B < K$. So the mBS can serve at most $B$ users simultaneously in a time slot. In each time slot, the mBS needs to select the $B$ users\footnote{If only $B^{\prime} < B$ users have non-empty queues in a time slot, then $B- B^{\prime}$ of the $B$ beams are deactivated in the slot.} to which beams will be assigned. In time slot $n$, the channel between the mBS and user $i \in \{1, \ldots, K\}$ is a Bernoulli random variable, say $S_n^i$, with success probability $d_i$. That is, $P(S_n^i = 1) = d_i$ and $P(S_n^i = 0) = 1- d_i$.  $S_n^i = 1$ (respectively, $S_n^i = 0$) denotes that the quality of the channel from the mBS to user $i$ is good (respectively, bad) in time slot $n$. Also, if user $i$ is assigned a beam in  slot $n$, and $S_n^i = 1$ (respectively, $S_n^i = 0$), then one packet (respectively, no packet) can be successfully sent from the mBS to user $i$ in the slot. Fig. \ref{fig:sys_model} illustrates the system model.

At the beginning of time slot $n$, let $X_n^i$ be the length of the queue at the mBS of packets to be sent to user $i$. Also, at the end of time slot $n$, $1$ packet arrives into the queue of user $i$ at the mBS with probability (w.p.) $a_i \in (0,1)$, and no packet arrives w.p. $1 - a_i$. The queue length changes over time as in the following equation:
\begin{equation} \label{state_update_equation}
    X_{n+1}^i = (X_n^i - S_n^i U_n^i)^+ + A_n^i,    
\end{equation}
where $x^+ := \max(0,x)$, and $U_n^i$ is a binary decision variable, which indicates whether a beam is formed from the mBS to user $i$ in time slot $n$ or not, and is given by:
\begin{equation*} 
U_n^i=\begin{cases}
        1, & \text{if a beam is assigned to user $i$ in slot $n$},\\ 
        0, & \text{else},
       \end{cases}
\end{equation*}
and:
\begin{equation*} 
\resizebox{246pt}{!}{$
A_n^i=\begin{cases}
        1, & \text{if a packet arrives into the queue of user $i$ in slot $n$},\\ 
        0, & \text{else}.
       \end{cases}$}
\end{equation*}

Let $H_i(X_n^i)$ be the packet holding cost at the mBS for user $i$ in time slot $n$. The function $H_i(\cdot)$ is assumed to be convex and non-decreasing in its argument. Also, if a beam is assigned to a user $i$ in a time slot, then a cost of $P_i > 0$ is incurred in the slot. This cost may be due to the energy consumed in forming a beam. Hence, the total cost experienced by the mBS in slot $n$ is $\sum_{i=1}^{K} \left[ H_i(X_n^i) + P_i U_n^i \right]$. Our goal is to devise a non-anticipating scheduling policy \cite{hordijk1983average}, which minimizes the long-run expected time-averaged cost experienced by the mBS, by appropriately selecting the control variables $U_n^i, i \in \{1, \ldots, K \}$, in every time slot $n$. In particular, the objective is as follows:
\begin{align} \label{primary_objective}
 \min & \underset {T \uparrow \infty} {\text{ limsup }} E \left[\frac{1}{T} \sum_{n=0}^{T-1}\sum_{i=1}^{K} \left\{ H_i \left (X_{n}^{i}\right) + P_i U_n^i \right\} \right], \notag\\
& \text{s.t.} \sum_{i=1}^{K} U_{n}^{i} = B, \forall n.
\end{align}
In each time slot $n$, the scheduling decisions $U_n^i, i \in \{1, \ldots, K \}$, need to be taken by the mBS \emph{before} observing the values of the variables $S_n^i, i \in \{1, \ldots, K\}$. In summary, the problem is to select $B$ users, to each of whom a beam is assigned in slot $n$, out of the $K$ users, so as to minimize the expected time-averaged holding cost. Note that minimizing the holding cost would lead to low average delays experienced by packets in the system.
\begin{remark}
    Note that if a beam is assigned to user $i$ in time slot $n$, then a packet sent from the mBS to user $i$ is successfully received at the user w.p. $d_i$. The parameter $d_i$ models path loss, blockage, channel fading, errors in transmission, etc.
\end{remark}

The constrained optimization problem \eqref{primary_objective} is provably hard to solve because of the exact per-stage constraint $\sum_{i=1}^{K} U_{n}^{i} = B$ \cite{papadimitriou1994complexity}. This hard constraint, which states that exactly $B$ beams must be assigned to $B$ users in each time slot, is relaxed using the approach proposed by Whittle \cite{whittle1988restless}. The relaxed constraint allows $B$ beams to be assigned to $B$ users \emph{on average} and is given by: 
\begin{equation} \label{relaxed_constraint}
\underset {T \uparrow \infty} {\text{ limsup }} \frac{1}{T} \sum_{n=0}^{T-1}\sum_{i=1}^{K} E \left [U_{n}^{i}\right ] = B.
\end{equation}
Next, a Lagrange multiplier strategy \cite{borkar2002convex} is used to convert the relaxed problem into an unconstrained problem, which is as follows:
\begin{align} \label{unconstrained_problem}
    \min \underset {T \uparrow \infty} {\text{ limsup }} \frac{1}{T} \sum_{n=0}^{T-1} &E \biggl [\sum_{i=1}^{K} \left\{ H_i \left (X_{n}^{i}\right) + P_i U_n^i \right\} \notag \\
    &+ \lambda \left ( B - \sum_{i=1}^{K} U_n^i \right ) \biggl],
\end{align}
where $\lambda$ is the Lagrange multiplier. Adding and subtracting the constant $\lambda(K-B)$ inside the argument of the expectation operator in \eqref{unconstrained_problem}, we get:
\begin{align} \label{simplified_term}
    \sum_{i=1}^{K} & \left \{H_i \left (X_{n}^{i}\right) + P_i U_n^i \right \} + \lambda \left ( B - \sum_{i=1}^{K} U_n^i \right ) \notag\\
    =& \sum_{i=1}^{K} \left \{ H_i \left (X_{n}^{i}\right)+ P_i U_n^i \right \}+ \lambda \left ( B - \sum_{i=1}^{K} U_n^i \right ) \notag \\
    &+ \lambda(K-B) - \lambda(K-B) \notag \\
    =& \sum_{i=1}^{K} \left \{H_i \left (X_{n}^{i}\right) + P_i U_n^i \right \} + \lambda \left ( K - \sum_{i=1}^{K} U_n^i \right )  \notag \\
    &- \lambda(K-B) \notag \\
    =& \sum_{i=1}^{K} \left \{H_i \left (X_{n}^{i}\right) + P_i U_n^i \right \}+ \lambda \sum_{i=1}^{K} \left ( 1 -  U_n^i \right ) - \lambda(K-B)\notag\\
    =& \sum_{i=1}^{K} \left \{ H_i \left (X_{n}^{i}\right) + P_i U_n^i + \lambda \left ( 1 -  U_n^i \right ) - \lambda \left( 1-\frac{B}{K} \right) \right \}.
\end{align}
For a given $\lambda$, the last term inside the braces in \eqref{simplified_term} is a constant and can be ignored during the minimization. Therefore, using \eqref{simplified_term}, the unconstrained problem \eqref{unconstrained_problem} becomes:
\begin{align} \label{final_unconstrained_problem}
    \min &\underset {T \uparrow \infty} {\text{ limsup }} \frac{1}{T} \sum_{n=0}^{T-1} \sum_{i=1}^{K} E \left [  H_i \left (X_{n}^{i}\right) + P_i U_n^i + \lambda \left ( 1 - U_n^i \right) \right].
\end{align}
As in the analysis by Whittle \cite{whittle1988restless}, $\lambda$ can be viewed as a penalty or a tax. When a beam from the mBS to a particular user $i$ is \emph{not} formed in time slot $n$, i.e., $U_n^i = 0$, then the penalty or tax is added to the holding cost, $H_i \left (X_{n}^{i}\right)$, experienced by the mBS due to the queue of packets to be sent to user $i$. For a given $\lambda$, the minimization problem \eqref{final_unconstrained_problem} decouples into individual controlled scheduling problems corresponding to different users. The controlled scheduling problem corresponding to user $i$ is a MDP, with state $X_n^i$ and action $U_n^i$, which is given by:
\begin{align} \label{final_objective_modified}
    \min &\underset {T \uparrow \infty} {\text{ limsup }} \frac{1}{T} \sum_{n=0}^{T-1}  E \left [  H_i \left (X_{n}^{i}\right) + P_i U_n^i + \lambda \left ( 1 - U_n^i \right ) \right], \notag \\ &\text{s.t.  } U_n^i \in \{0,1\}, \; \forall n.
\end{align}
The original problem \eqref{primary_objective} is said to be Whittle indexable \cite{whittle1988restless}, if for the decoupled MDP \eqref{final_objective_modified} corresponding to each user $i$, for all possible values of the parameters $(P_i, d_i, a_i) \in (0,\infty) \times (0,1) \times (0,1)$, and for all possible convex non-decreasing functions $H_i(\cdot)$, the set of states  known as ``passive states" decreases monotonically from the entire state space to the empty set as $\lambda$ increases from $-\infty$ to $\infty$. Here, the passive states are those in which a beam from the mBS to user $i$ is not formed. Conversely, the states in which a beam from the mBS to user $i$ is formed are termed as ``active states". For a given state, the value of $\lambda$ at which the controller is indifferent between the assignment and non-assignment of a beam to the user, is called the ``Whittle index". The proposed Whittle index based policy operates as follows. In each time slot $n$, we calculate the Whittle index  of each user $i \in \{1, \ldots, K\}$. The $B$ users with the $B$ smallest values of the Whittle index are assigned beams, and the other $K-B$ users are not assigned beams in the time slot. 

\begin{remark}
    The system model described above is particularly well-suited for mmWave networks since typically, communication takes place using narrow beams between the mBS and users in such networks. However, all our results are also applicable to other wireless networks, such as sub-6 GHz wireless networks, in which communication via narrow beams is used.
\end{remark}

\section{Dynamic Programming Equation} \label{value_Function_sec}
As mentioned earlier, for a given $\lambda$, the minimization problem \eqref{final_unconstrained_problem} separates out into individual controlled scheduling problems, given in \eqref{final_objective_modified}, corresponding to different users $i$. As the proof of Whittle indexability for the individual controlled scheduling problems is the same for every user $i$, henceforth, we omit the index $i$ from all notation. Let $X_{n} = x$ denote the current state of user $i$, i.e., the queue length of packets to be sent to user $i$ from the mBS. We claim that the Dynamic Programming Equation (DPE) for the individual MDP of user $i$ in \eqref{final_objective_modified} is given by:
\begin{align}\label{Value_Function}
    V(x) =& H(x)-\eta + \min\biggl[ V(x) a d + V(x) (1-a)(1-d) \notag\\ 
          &+ V(x+1) a (1-d) + V(x-1)(1-a)d + P;\notag\\
          &\lambda + V(x) (1-a)+V(x+1) a \biggr],
\end{align}
where $V(\cdot)$ is the value function for the MDP in \eqref{final_objective_modified} and $\eta$ is a constant, whose value will be specified later. The first (respectively, second) input of the $\min$ in the RHS of \eqref{Value_Function} corresponds to the control $U = 1$ (respectively, $U = 0$). In the rest of this section, we prove \eqref{Value_Function}. 

For any stationary policy, say $\pi$, for the MDP in \eqref{final_objective_modified}, the discounted cost over an infinite horizon for the controlled process with a discount factor $\gamma \in (0,1)$ and the initial state $x$ can be expressed as follows:
\begin{align*}
    C^{\gamma}(x, \pi) = E \biggl [  \sum_{n=0}^{\infty} &\gamma^n \biggl ( H \left (X_n \right) + P U_n \\
    &+ \lambda(1-U_n) \biggl )|X_0 = x \biggl ].
\end{align*}
The value function of the discounted problem is the minimum over all stationary control policies $\pi$ and is given by:
\begin{equation*}
    V^{\gamma}(x) = \min_{\pi} C^{\gamma}(x,\pi).
\end{equation*}
If the transition probability function of the Discrete-Time Markov Chain (DTMC) induced by a control $U$ for the MDP in \eqref{final_objective_modified} is denoted as $p_{.|.}(U)$, then the value function $V^{\gamma}(x)$ for the current state $x$ can be characterized by the following DPE:
\begin{align} \label{EQ:discounted:problem:DPE}
    V^{\gamma}(x) =& \min_{U \in \{0,1\}} \biggl [ H(x) + P U + \lambda(1-U) \notag \\
    &+ \gamma\sum_{j}^{}p_{j|x}(U)V^{\gamma}(j) \biggl ].
\end{align}
Let $\bar{V}^{\gamma}(\cdot) = V^{\gamma}(\cdot) - V^{\gamma}(0)$. The DPE in \eqref{EQ:discounted:problem:DPE} can be written as follows:
\begin{equation} \label{beta_val_fun_bar}
\begin{split}
    \bar{V}^{\gamma}(x) = &\min_{U \in \{0,1\}} \biggl [  H(x) + P U + \lambda(1-U)  - (1-\gamma)V^{\gamma}(0)  \\
 &+ \gamma\sum_{j}^{}p_{j|x}(U)\bar{V}^{\gamma}(j) \biggl ].
\end{split}
\end{equation}
The value function $V(\cdot)$ and constant $\eta$ in \eqref{Value_Function} can be derived from \eqref{beta_val_fun_bar} using the following lemma.

\begin{lemma}  \label{lem:lemma1}
The quantities $V(\cdot)$ and $\eta$ satisfying \eqref{Value_Function} can be derived as: $\lim_{\gamma \uparrow 1} \bar{V}^{\gamma}(\cdot) = V(\cdot)$ and $\lim_{\gamma \uparrow 1} (1-\gamma) V^{\gamma}(0) = \eta$. The constant $\eta$ in \eqref{Value_Function} is unique and is equal to the optimal long-run expected average cost of the individual MDP in \eqref{final_objective_modified}. Under an optimal policy and the additional constraint $V(0) = 0$, $V(\cdot)$ also maintains uniqueness in states that are positive recurrent. For a given state $x$, the optimal choice of $U$ is obtained by finding the argmin of the RHS of \eqref{Value_Function}.     
\end{lemma}
\begin{IEEEproof}
The proof of this lemma is analogous to that of Lemma 4 in \cite{borkar2022whittle} and is omitted for brevity.
\end{IEEEproof} 

\section{Structural Properties of Value Function} \label{Structural_prop}
We prove some structural properties of the value function in this section. Later, these structural properties will be used to prove that the optimal policy is a threshold policy and to show that the beam scheduling problem is Whittle indexable.
\begin{lemma} \label{lem:lemma2}
  The function $V(\cdot)$ in \eqref{Value_Function} is a non-decreasing function.
\end{lemma}
\begin{IEEEproof}
  To prove this lemma, we use induction. For $w \geq 1$, the finite horizon DPE with a discount factor $\gamma$ and current state $x$ is given by:
\begin{align}\label{Finite_Hor_Discounted}
{V}_w^{\gamma}(x) =& \min_{U \in \{0,1\}} \biggl[ H(x)+ PU + \lambda(1-U)+ \gamma \biggl\{ V_{w-1}^{\gamma}(x) \times  \notag\\
&(1-d)(1-a) + V_{w-1}^{\gamma}(x+1)(1-d)a   \notag \\
&+ V_{w-1}^{\gamma}(x-U)d (1-a) \notag \\
&+ V_{w-1}^{\gamma}(x-U+1)da \biggr \} \biggr].
\end{align}
Here, $V_{0}^{\gamma}(x)=H(x), \forall x \geq 0$. Since $H(\cdot)$ is non-decreasing, $V_{0}^{\gamma}(x_1) \geq V_{0}^{\gamma}(x_2)$ for all $x_1, x_2$ s.t. $x_1 > x_2 \geq 0$. Assume that:
\begin{equation} \label{assumption1}
    V_{w-1}^{\gamma}(x_1) \geq V_{w-1}^{\gamma}(x_2), \; \forall x_1, x_2 \text{ s.t. } x_1 > x_2 \geq 0.
\end{equation}
We need to prove that $V_{w}^{\gamma}(x_1) \geq V_{w}^{\gamma}(x_2), \forall x_1, x_2$ s.t. $x_1 > x_2 \geq 0$. Let the minimum value of the RHS of \eqref{Finite_Hor_Discounted} at current state $x_1$ (respectively, $x_2$) be achieved at $U_1$ (respectively, $U_2$). Then using \eqref{Finite_Hor_Discounted}, we can write:
\begin{subequations}\label{l2_subequations}
\begin{align}
    V_{w}^{\gamma}(x_{1}) =& H(x_1)+ PU_1 + \lambda(1-U_1)+ \gamma \biggl\{ V_{w-1}^{\gamma}(x_1) \times \notag\\
    &(1-d)(1-a)+ V_{w-1}^{\gamma}(x_1+1)(1-d)a   \notag \\
    &+ V_{w-1}^{\gamma}(x_1-U_1)d (1-a) \notag \\
    &+ V_{w-1}^{\gamma}(x_1-U_1+1)da \biggr \},  \label{l2_sub_eq1}\\
    V_{w}^{\gamma}(x_{2}) =& H(x_2)+ P U_2 + \lambda(1-U_2)+ \gamma \biggl\{ V_{w-1}^{\gamma}(x_2) \times \notag\\
    &(1-d)(1-a)+ V_{w-1}^{\gamma}(x_2+1)(1-d)a \notag \\
    &+ V_{w-1}^{\gamma}(x_2-U_2)d (1-a) \notag \\
    &+ V_{w-1}^{\gamma}(x_2-U_2+1)da \biggr \}. \label{l2_sub_eq2}
\end{align}
\end{subequations}
Taking the difference between \eqref{l2_sub_eq1} and \eqref{l2_sub_eq2} gives:
\begin{align} \label{l2_diff_eq}
    V_{w}^{\gamma}&(x_{1}) - V_{w}^{\gamma}(x_{2}) \notag \\
    =& H(x_1)-H(x_2) + P(U_1-U_2)+(U_{2}-U_{1})\lambda \notag \\
    &+ \gamma \biggl[\biggl( V_{w-1}^{\gamma}(x_1) -V_{w-1}^{\gamma}(x_2)\biggr)(1-d)(1-a)\biggr] \notag \\
    &+ \gamma \biggl[ \biggl( V_{w-1}^{\gamma}(x_1+1) -  V_{w-1}^{\gamma}(x_2+1)\biggr)(1-d)a  \biggr] \notag \\
    &+ \gamma \biggl[\biggl(V_{w-1}^{\gamma}(x_1-U_1)-V_{w-1}^{\gamma}(x_2-U_2)\biggr)d(1-a) \biggr] \notag \\
    &+ \gamma \biggl[ \biggl( V_{w-1}^{\gamma}(x_1-U_1+1)- V_{w-1}^{\gamma}(x_2-U_2+1)\biggr)da   \biggr].  
\end{align}
The controls $U_1$ and $U_2$ are binary, i.e., they take the value $0$ or $1$. Consider the case where $U_{1} = U_{2} =U$. In this case, \eqref{l2_diff_eq} becomes:
\begin{align}\label{inceasingfunc_final}
V_{w}^{\gamma}&(x_{1}) - V_{w}^{\gamma}(x_{2}) \notag \\
    =& H(x_1)-H(x_2)  \notag \\
    &+ \gamma \biggl[\biggl( V_{w-1}^{\gamma}(x_1) -V_{w-1}^{\gamma}(x_2)\biggr)(1-d)(1-a)\biggr] \notag\\
    &+ \gamma \biggl[ \biggl( V_{w-1}^{\gamma}(x_1+1) -  V_{w-1}^{\gamma}(x_2+1)\biggr)(1-d)a  \biggr] \notag\\
    &+ \gamma \biggl[\biggl(V_{w-1}^{\gamma}(x_1-U)-V_{w-1}^{\gamma}(x_2-U)\biggr)d(1-a)  \biggr] \notag\\
    &+ \gamma \biggl[ \biggl( V_{w-1}^{\gamma}(x_1-U+1)- V_{w-1}^{\gamma}(x_2-U+1)\biggr)da   \biggr].  
\end{align}
From the non-decreasing nature of $H(\cdot)$ and from \eqref{assumption1}, it follows that the quantity in \eqref{inceasingfunc_final} is non-negative. Now, consider the case where $U_{1} \neq U_{2}$. Suppose $U_{1} = 1, U_{2} = 0$, i.e., the minimum value of $V_{w}^{\gamma}(x_{1})$ is achieved at $U_{1} =1 $ and that of $V_{w}^{\gamma}(x_{2})$ is achieved at $U_{2} = 0$. Then, from the above results for the case where $U_1 = U_2$, it follows that:
\begin{align*}
    V_{w}^{\gamma}(x_1)\biggl|_{U_{1}=1} \geq V_{w}^{\gamma}(x_2)\biggl|_{U_{2}=1} \geq V_{w}^{\gamma}(x_2)\biggl|_{U_{2}=0} .
\end{align*}
Therefore, $V_{w}^{\gamma}(x_1) \geq V_{w}^{\gamma}(x_2)$. If $U_{1} = 0, U_{2} = 1$, similar to the above, we can write:
\begin{align*}
    V_{w}^{\gamma}(x_1)\biggl|_{U_{1}=0} \geq V_{w}^{\gamma}(x_2)\biggl|_{U_{2}=0} \geq V_{w}^{\gamma}(x_2)\biggl|_{U_{2}=1}  .
\end{align*}
Therefore, $V_{w}^{\gamma}(x_1) \geq V_{w}^{\gamma}(x_2), \forall \; U_1, U_2$. Taking limits as $w \uparrow \infty$, we get the following inequality  for the infinite horizon: $V^{\gamma}(x_1) \geq V^{\gamma}(x_2)$. Further, by taking limits as $\gamma \uparrow 1$, we can conclude that:
\begin{equation*}
    V(x_1) \geq V(x_2) \; \forall x_1, x_2 \text{ s.t. } \; x_1 > x_2 \geq  0.
\end{equation*}
The result follows.
\end{IEEEproof}

\begin{lemma} \label{lem:lemma3}
The function $V(\cdot)$ in \eqref{Value_Function} has the non-decreasing differences property, i.e., if $b>0$ is an integer and  $x_{1}$ and $x_{2}$ are non-negative integers s.t. $x_1 > x_2$, then:
\begin{equation*}
    V(x_{1}+b) -  V(x_{1}) \geq  V(x_{2}+b) -  V(x_{2}).
\end{equation*}
\end{lemma}
\begin{IEEEproof}
Consider the DPE in \eqref{Finite_Hor_Discounted}. For states $x_1$ and $x_2$ s.t. $x_{1}>x_{2}$ and $b>0$, let us prove the following inequality using induction:
\begin{equation}\label{Increasing_Diff}
    V_{w}^{\gamma}\left(x_{1}+b\right)-V_{w}^{\gamma}\left(x_{1}\right)-V_{w}^{\gamma}\left(x_{2}+b\right)+V_{w}^{\gamma}\left(x_{2}\right) \geq 0.
\end{equation}
The inequality in \eqref{Increasing_Diff} is clearly true for $w=0$ due to the convex non-decreasing nature of $H(\cdot)$; assume that it is also true for $w - 1$. That is: 
\begin{equation}\label{Assumption_1}
    V_{w-1}^{\gamma}(x_{1}+b) -  V_{w-1}^{\gamma}(x_{1}) -  V_{w-1}^{\gamma}(x_{2}+b) +  V_{w-1}^{\gamma}(x_{2}) \geq 0.
\end{equation}
Let us define $I^{1} = V_{w}^{\gamma}(x_{1}+b) -  V_{w}^{\gamma}(x_{1}) $ and $I^{2} = V_{w}^{\gamma}(x_{2}+b) -  V_{w}^{\gamma}(x_{2}) $. Our aim is to prove that $I^{1} - I^{2} \geq 0$. Suppose the minimum values of $V_{w}^{\gamma}(\cdot)$ in the RHS of \eqref{Finite_Hor_Discounted} at the states $x_{1}+b, x_{1}, x_{2}+b$, and $x_{2}$ are achieved at the binary controls $U_{1}, U_{2}, U_{3}$, and $U_{4}$, respectively. From \eqref{Finite_Hor_Discounted}, we can write:
\begin{align}\label{Difference_Equation}
    I^{1}&-I^{2} \notag \\ 
    =& \biggl \{H(x_1+b)-H(x_1)-H(x_2 +b) + H(x_2) \biggl \}\notag \\
    &+ P(U_1 - U_2 -U_3 + U_4) + \lambda \biggl(U_{2}-U_{1} -U_{4} +U_{3}\biggr) \notag \\
    & + \gamma \biggl( V_{w-1}^{\gamma}(x_1+b) -  V_{w-1}^{\gamma}(x_1) - V_{w-1}^{\gamma}(x_2+b)\notag\\
    &  + V_{w-1}^{\gamma}(x_2)\biggr) (1-d)(1-a)  + \gamma \biggl(V_{w-1}^{\gamma}(x_{1} +1 + b) \notag\\
    & -V_{w-1}^{\gamma}(x_{1} + 1) - V_{w-1}^{\gamma}(x_{2}+ 1 + b) \notag\\
    &+ V_{w-1}^{\gamma}(x_{2} +1  ) \biggr) (1-d)a + \gamma \biggl(V_{w-1}^{\gamma}(x_1+b-U_{1})\notag\\
    & -V_{w-1}^{\gamma}(x_1-U_{2}) - V_{w-1}^{\gamma}(x_2+b-U_{3}) \notag\\
    & + V_{w-1}^{\gamma}(x_2-U_{4})\biggr)d(1-a) + \gamma \biggl( V_{w-1}^{\gamma}(x_1+b-U_{1} \notag \\
    &+1) - V_{w-1}^{\gamma}(x_1-U_{2}+1) - V_{w-1}^{\gamma}(x_2+b-U_{3}+1) \notag\\
    & + V_{w-1}^{\gamma}(x_2-U_{4}+1)\biggr)da.
\end{align}
The first term in the RHS of \eqref{Difference_Equation} (enclosed in curly braces) is always non-negative due to the convexity of $H(\cdot)$. The fourth and fifth terms of the RHS of \eqref{Difference_Equation} are always non-negative as they are independent of the controls and due to \eqref{Assumption_1}. Denote the sum of the remaining terms of the RHS of \eqref{Difference_Equation} by $T$. That is:
\begin{align} \label{Separate_term}
    T =& P(U_1 - U_2 -U_3 + U4) + \lambda \biggl(U_{2}-U_{1}-U_{4}+U_{3}\biggr) \notag\\
    &+ \gamma \biggl(V_{w-1}^{\gamma}(x_1+b-U_{1}) -V_{w-1}^{\gamma}(x_1-U_{2}) \notag \\
    & - V_{w-1}^{\gamma}(x_2+b-U_{3}) + V_{w-1}^{\gamma}(x_2-U_{4})\biggr)d(1-a) \notag \\
    & + \gamma \biggl( V_{w-1}^{\gamma}(x_1+b-U_{1} +1) - V_{w-1}^{\gamma}(x_1-U_{2}+1) \notag\\
    & - V_{w-1}^{\gamma}(x_2+b-U_{3}+1)+ V_{w-1}^{\gamma}(x_2-U_{4}+1)\biggr)da.
\end{align}
We need to prove that $T \geq 0$. First, we consider the case in which  the minimum values of $V_{w}^{\gamma}(\cdot)$ in the RHS of \eqref{Finite_Hor_Discounted} at the states  $x_{1}+b$ and $x_{2}+b$ are achieved at the same binary control, i.e., $U_{1} = U_{3} = U^{1}$ (say). Similarly, assume that the minimum values of $V_{w}^{\gamma}(\cdot)$ in the RHS of \eqref{Finite_Hor_Discounted} at the states  $x_{1}$ and $x_{2}$ are achieved at the same binary control, i.e., $U_{2} = U_{4} = U^{2}$ (say). Then \eqref{Separate_term} can be written as:
\begin{align} \label{second_separate}
    T = &\gamma \biggl(V_{w-1}^{\gamma}(x_1+b-U^{1}) -V_{w-1}^{\gamma}(x_1-U^{2}) \notag\\
    &- V_{w-1}^{\gamma}(x_2+b-U^{1}) + V_{w-1}^{\gamma}(x_2-U^{2})\biggr)d(1-a) \notag \\
    &+ \gamma \biggl( V_{w-1}^{\gamma}(x_1+b-U^{1}+1) - V_{w-1}^{\gamma}(x_1-U^{2}+1) \notag\\
    & - V_{w-1}^{\gamma}(x_2+b-U^{1}+1) + V_{w-1}^{\gamma}(x_2-U^{2}+1)\biggr)da.
\end{align} 
If $U^1 = U^2$, then from \eqref{Assumption_1}, it follows that $T \geq 0$. Let us show the non-negativity of $T$ in the cases in which the values of $U^1$ and $U^2$ are different. Suppose $U^1 = 1$ and $U^2=0$. Then \eqref{second_separate} becomes:
\begin{align*}
    T = &\gamma \biggl(V_{w-1}^{\gamma}(x_1+b-1) -V_{w-1}^{\gamma}(x_1) - V_{w-1}^{\gamma}(x_2+b-1)\notag\\
    &+ V_{w-1}^{\gamma}(x_2)\biggr)d(1-a) + \gamma \biggl( V_{w-1}^{\gamma}(x_1+b) - V_{w-1}^{\gamma}(x_1  \notag\\
    &+1)- V_{w-1}^{\gamma}(x_2+b) + V_{w-1}^{\gamma}(x_2+1)\biggr)da.   
\end{align*}
From \eqref{Assumption_1} and the fact that $b > 0$, it follows that $T \geq 0$. Similarly, if $U^1 = 0$ and $U^2=1$, then \eqref{second_separate} becomes: 
\begin{align*}
    T = &\gamma \biggl(V_{w-1}^{\gamma}(x_1+b) -V_{w-1}^{\gamma}(x_1-1) - V_{w-1}^{\gamma}(x_2+b) \notag\\
    & + V_{w-1}^{\gamma}(x_2-1)\biggr)d(1-a) + \gamma \biggl( V_{w-1}^{\gamma}(x_1+b+1)\notag \\
    & - V_{w-1}^{\gamma}(x_1) - V_{w-1}^{\gamma}(x_2+b+1) + V_{w-1}^{\gamma}(x_2)\biggr)da.    
\end{align*}
Again, from \eqref{Assumption_1} and the fact that $b>0$, it follows that $T \geq 0$. Thus, in all possible cases under the assumption that $U_{1}=U_{3}$ and $U_{2} = U_{4}$ (where $U_{1}, U_{2}, U_{3}$, and $U_{4}$ are the controls at which the minimum values of $V_{w}^{\gamma} (\cdot)$ in the RHS of \eqref{Finite_Hor_Discounted} at the states $x_{1}+b, x_{1}, x_{2}+b$ and $x_{2}$, respectively, are achieved), we have proved that the expression in \eqref{Difference_Equation} is non-negative. Specifically, we have proved that $\left.I^{1}\right|_{U_{1}=m, U_{2}=q} \geq\left. I^{2}\right|_{U_{3}=m, U_{4}=q}$,  $\forall  m, q \in\{0,1\}$. Now suppose $m=0$ and $q=1$; then we get:
\begin{equation*}
    \left.I^{1}\right|_{U_{1}=0, U_{2}=1} \geq\left. I^{2}\right|_{U_{3}=0, U_{4}=1}.
\end{equation*}
If the minimum values of $V_{w}^{\gamma}(\cdot)$ in the RHS of \eqref{Finite_Hor_Discounted} at the states $x_{1}+b$ and $x_{1}$ are attained at $U_{1}=0$ and $U_{2}=1$, respectively, then the minimum of $I^{1}$ is achieved when $V_{w}^{\gamma}(x_{1}+b)$ is at its minimum and $V_{w}^{\gamma}(x_{1})$ is at its maximum, i.e., at $U_{1}=0$ and $U_{2}=0$. Similarly, the maximum of $I^{1}$ is attained at $U_{1}=1$ and $U_{2}=1$. Thus, we can write:
\begin{equation}\label{I1equation}
\resizebox{245pt}{!}{$
    \left.I^{1}\right|_{U_{1}=1, U_{2}=1} \geq\left. I^{1}\right|_{U_{1}=0, U_{2}=1}, \left. I^{1}\right|_{U_{1}=1, U_{2}=0} \geq\left. I^{1}\right|_{U_{1}=0, U_{2}=0}. $}
\end{equation}
If the minimum values of $V_{w}^{\gamma}(\cdot)$  in the RHS of \eqref{Finite_Hor_Discounted} at the states $x_{2}+b$ and $x_{2}$ are attained at $U_{3}=1$ and $U_{4}=0$, respectively, then the minimum of $I^{2}$ is achieved when $V_{w}^{\gamma}(x_{2}+b)$ is at its minimum and $V_{w}^{\gamma}(x_{2})$ is at its maximum, i.e., at $U_{3}=1$ and $U_{4}=1$. Similarly, the maximum of $I^{2}$ is attained at $U_{3}=0$ and $U_{4}=0$. Thus, we can write:
\begin{equation}\label{I2equation}
\resizebox{245pt}{!}{$
    \left.I^{2}\right|_{U_{3}=0, U_{4}=0} \geq\left. I^{2}\right|_{U_{3}=0, U_{4}=1},\left. I^{2}\right|_{U_{3}=1, U_{4}=0} \geq\left. I^{2}\right|_{U_{3}=1, U_{4}=1}. $}
\end{equation}
However, we know that $\left.I^{1}\right|_{U_{1}=0, U_{2}=0} \geq\left. I^{2}\right|_{U_{3}=0, U_{4}=0}$. So from \eqref{I1equation} and \eqref{I2equation}, we can conclude that
$\left.I^{1}\right|_{U_{1}=0, U_{2}=1} \geq\left. I^{2}\right|_{U_{3}=1, U_{4}=0}$. Similarly, we can show that $\left.I^{1}\right|_{U_{1}, U_{2}} \geq\left. I^{2}\right|_{U_{3}, U_{4}}$ for all possible values of $U_{1}$, $U_{2}$, $U_{3}$, and $U_{4}$. This completes the proof of \eqref{Increasing_Diff}. 

Next, taking limits as $w \uparrow \infty$ in \eqref{Increasing_Diff}, we get that the infinite-horizon value function, $V^{\gamma}(\cdot)$, also satisfies the non-decreasing differences property;  again, by taking limits as $\gamma \uparrow 1$, we can conclude that the value function $V(\cdot)$ has non-decreasing differences. The result follows.
\end{IEEEproof}

\section{Threshold Behavior of Optimal Policy} \label{Section7_thres}
In this section, we show that the optimal policy for the MDP \eqref{final_objective_modified} is of threshold type. That is, under the optimal policy, there exists a state, say $t$, which serves as a threshold, determining whether a beam is assigned to the user or not for a given value of $\lambda$. For states at or below this threshold, i.e., states $x \leq t$, a beam is not assigned to the user and hence there are no packet departures, whereas, for states above the threshold, i.e., states $x > t$, a beam is assigned to the user and hence there can be packet departures under the optimal policy. The value of this threshold state, $t$, is determined by $\lambda$. Fig. \ref{fig:MDP_model} illustrates the threshold behavior of the optimal policy graphically. The following lemma proves the threshold nature of the optimal policy.
\begin{lemma} \label{lem:lemma5}
The optimal policy is a threshold policy.    
\end{lemma}
\begin{IEEEproof}
For a state $x$, if the optimal action is the assignment of a beam, then to prove this lemma, we have to show that for states above $x$ as well, the optimal action is the assignment of a beam. Consider the function $g(x) = \mathbb{E}[V\left((x-D)^+ +A\right)] - \mathbb{E}[V(x+A)]$, where $x$ represents the number of packets in the queue of packets to be sent to a given user from the mBS, $D$ is the number of packet departures from the queue, and $A$ is the number of packets that arrive in the queue in a given time slot. It follows from the definition of threshold policy that an optimal policy is a threshold policy if $g(x+w) \leq g(x)$ for $w>0$ \cite{borkar2022whittle}. Hence, proving that $g(x+1)-g(x) \leq 0$ is sufficient. We prove this in the rest of this proof. We have: 
\begin{figure}
\centerline{\includegraphics[width=\linewidth]{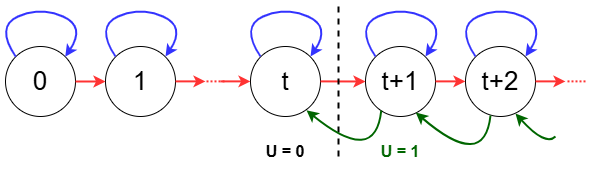}}
\caption{The figure shows that for every state greater than $t$, a beam is allocated to the user and hence a packet departure can take place, while for every state below or equal to $t$, a beam is not allocated to the user, resulting in no packet departure.}
\label{fig:MDP_model}
\end{figure}
\begin{align} \label{exp_eq_dep}
    \mathbb{E}&\left[V\left((x-D)^+ +A\right)\right] \notag \\
    =& V(x+1) a (1-d) +  V(x-1)(1-a)d + V(x) a d\notag \\
    & + V(x) (1-a)(1-d). 
\end{align}
Similarly:
\begin{align} \label{exp_eq}
    \mathbb{E}\left[V(x+A)\right] =& V(x) (1-a)+V(x+1) a.
\end{align}
Therefore, from \eqref{exp_eq_dep} and \eqref{exp_eq}, we get:
\begin{align} \label{gx_eq}
g(x)& \notag \\
 =& \mathbb{E}\left[V\left((x-D)^+ +A\right)\right] - \mathbb{E}\left[V\left(x+A\right)\right] \notag\\
 =& -V(x+1) a d + V(x) (2a d -d) +  V(x-1)(1-a)d \notag \\ 
 =&- \biggl [ V(x+1) - 2V(x) + V(x-1)\biggr] ad - \biggl [ V(x) \notag \\
 & - V(x-1)\biggr]d.
 \end{align}
Replacing $x$ with $x+1$ in \eqref{gx_eq}, we get:
 \begin{align} \label{gx1_eq}
g(x+1) =& - \biggl [ V(x+2) - 2V(x+1) + V(x)\biggr] ad \notag \\
 &  - \biggl [ V(x+1)- V(x)\biggr]d.
 \end{align}
Subtracting \eqref{gx_eq} from \eqref{gx1_eq} we get:
\begin{align} \label{diff_eq_g}
    g(x+&1) - g(x) \notag \\
    =& -\biggl[\biggl\{ V(x+2) -2V(x+1) +V(x) \biggr \} \notag \\
    &-\biggl\{  V(x+1) - 2V(x) + V(x-1)  \biggr\}\biggr] ad \notag \\
    &- \biggl[  V(x+1)- 2V(x) + V(x-1) \biggr]d \notag \\
    =&-\biggl[ V(x+2) -2V(x+1) +V(x) \biggr ]ad \notag \\
    &+ \biggl[  V(x+1)- 2V(x) + V(x-1) \biggr](a-1)d.
\end{align}
In Lemma \ref{lem:lemma3}, if we substitute $x_1 = x+1, x_2=x$, and $b=1$, then we get $V(x+2)-2V(x+1)+V(x) \geq 0$. Thus, the first term in \eqref{diff_eq_g} is non-positive. Similarly, in Lemma \ref{lem:lemma3}, if we substitute $x_1 = x, x_2=x-1$, and $b=1$, then we get $V(x+1)-2V(x)+V(x-1) \geq 0$. As the value of the parameter $a \in (0,1)$, the term $(a-1)$ is always negative. Thus, the second term in  \eqref{diff_eq_g} is also non-positive. This implies that the entire expression in  \eqref{diff_eq_g} is always less than or equal to zero; so $g(x+1)-g(x) \leq 0$. It follows that the optimal policy is a threshold policy.
\end{IEEEproof}

Recall that a stationary control policy for the MDP \eqref{final_objective_modified} induces a DTMC. The following lemma proves a useful result about the stationary distribution of the DTMC induced under a threshold policy, which we will use later. 
\begin{lemma} \label{lem:lemma6}
    Let $v_t (\cdot)$ be the stationary distribution of the DTMC induced under the threshold policy with threshold $t$. Then $\sum_{q=0}^{t} v_t(q)$ is an increasing function of $t$.
\end{lemma}
\begin{IEEEproof}  
This result can be proved by using Lemma \ref{lem:lemma5} and following steps similar to those in the proof of Lemma 8 in \cite{borkar2022whittle}. We omit the details for brevity. 
\end{IEEEproof}

\section{Whittle Indexability} \label{Section8_wi}
Whittle indexability of the problem requires that as the value of $\lambda$ decreases from $\infty$ to $-\infty$, the set of passive states (i.e., the states for which a beam from the mBS to the user is not formed) monotonically increases from the empty set to the set of all possible states. To establish Whittle indexability, we need to introduce some supporting lemmas.

\begin{lemma} \label{lem:lemma7}
    Let $f: \mathbb{R} \times \mathbb{N} \rightarrow \mathbb{R}$ be supermodular, i.e., $\forall \lambda_1 > \lambda_2$ and $x_1 > x_2$:
    \begin{equation*}
        f(\lambda_2, x_1) + f(\lambda_1, x_2) \leq f(\lambda_1, x_1) + f(\lambda_2, x_2).
    \end{equation*}
Also, let $x(\lambda) \coloneqq \inf \{x^* : f(\lambda, x^*) \leq f(\lambda, x), \; \forall x\}$. Then $x(\lambda)$ is a non-increasing function of $\lambda$.
\end{lemma}
\begin{IEEEproof} 
       The proof of this result is similar to that of a result on submodularity presented in Section 10.2 of \cite{sundaram1996first}. We omit the proof for brevity.
\end{IEEEproof}

\begin{lemma} \label{lem:lemma8}
For the MDP \eqref{final_objective_modified}, let the average cost of the threshold policy with threshold $t$, under the tax $\lambda$, be:
    \begin{equation*}
       \mathfrak{f}(\lambda, t) = \sum_{j=0}^{\infty} \; H(j) v_{t} (j) + \lambda \sum_{j=0}^{t} v_{t}(j) + P \sum_{j=t+1}^{\infty} v_{t}(j).
    \end{equation*}
    Then the function $\mathfrak{f}$ is supermodular. 
\end{lemma}
\begin{IEEEproof} 
 To prove this, we need to show that $\forall \lambda_2 < \lambda_1$ and $t_2 < t_1$:
\begin{equation*}
    \mathfrak{f}(\lambda_1, t_2) + \mathfrak{f}(\lambda_2, t_1) \leq \mathfrak{f}(\lambda_1, t_1) + \mathfrak{f}(\lambda_2, t_2).
\end{equation*}
Substituting the value of $\mathfrak{f}(\lambda,t)$ and simplifying, the preceding inequality becomes:
\begin{align*}
    \lambda_1 \sum_{j=0}^{t_2} v_{t_2}(j) + \lambda_2 \sum_{j=0}^{t_1} v_{t_1}(j) \leq \lambda_1 \sum_{j=0}^{t_1} v_{t_1}(j) + \lambda_2 \sum_{j=0}^{t_2} v_{t_2}(j).
\end{align*}
As $\lambda_1 > \lambda_2$, the above inequality is equivalent to:
\begin{align*}
    \sum_{j=0}^{t_2} v_{t_2}(j) &\leq \sum_{j=0}^{t_1} v_{t_1}(j).
\end{align*}
This inequality follows from Lemma \ref{lem:lemma6}. Thus, $\mathfrak{f}$ is supermodular.
\end{IEEEproof}

\begin{theorem}
    The problem is Whittle indexable.
\end{theorem}
\begin{IEEEproof}
Under the unichain property \cite{sheldonross}, for any given stationary policy, say $\pi$, there exists a unique stationary distribution, denoted by $v$. Let $Z$ represent the set of passive states, i.e., states at which a beam from the mBS to the user is not formed, under the stationary policy $\pi$ with a given $\lambda$. If $Z'$ is the set of active states, then the optimal expected average cost under any stationary policy can be expressed as:
\begin{equation*}
    \eta(\lambda) = \inf_{\pi} \left \{ \sum_{j=0}^{\infty} H(j) v(j) + \lambda\sum_{j \in \emph{Z}}^{}v(j) +P \sum_{j \in \emph{Z'}} v(j) \right \}.
\end{equation*}
By Lemma \ref{lem:lemma5}, $\eta(\lambda)$ is the infimum over all threshold policies. Therefore, $\eta(\lambda)$ can be expressed as:
\begin{align*}
    \eta(\lambda) =& \sum_{j=0}^{\infty} H(j) v_{t(\lambda)}(j) + \lambda \sum_{j=0}^{t(\lambda)} v_{t(\lambda)}(j) \\
    & + P \sum_{j=t(\lambda)+1}^{\infty} v_{t(\lambda)}(j)\\
    =& \mathfrak{f}(\lambda, t(\lambda)),
\end{align*}
where $t(\lambda)$ is the optimal threshold for the given tax $\lambda$. By Lemma \ref{lem:lemma8}, $\mathfrak{f}$ is supermodular. So by Lemma \ref{lem:lemma7}, the threshold $t(\lambda)$ is a non-increasing function of $\lambda$. Under the optimal threshold policy, the set of passive states, $Z$, is $Z = \{0,1, \ldots, t(\lambda)\}$. Therefore, as \( \lambda \) increases from \(-\infty \) to \( +\infty \), \( Z \) monotonically decreases from the whole state space to the empty set. This proves the Whittle indexability of the problem.
\end{IEEEproof}

\section{Whittle Index Computation}\label{Section9_comp}
For a state $x$, the Whittle index $\lambda$ is computed iteratively. Specifically, with $\tau \geq 0$ being the iteration number, the value of $\lambda$ is updated according to the following equation:
\begin{equation} \label{update_lambda}
\resizebox{245pt}{!}{$
    \lambda_{\tau+1} = \lambda_{\tau} + \beta \biggl ( \sum_{j}^{}p_{dp}(j|x)V_{\lambda_{\tau}}(j) + P - \sum_{j}^{}p_{dn}(j|x)V_{\lambda_{\tau}}(j) - \lambda_{\tau} \biggl ),$}
\end{equation}
where $\beta >0$ represents a small step size, $V_{\lambda_{\tau}}(\cdot)$ is the value function, which appears in the DPE \eqref{Value_Function}, for the tax $\lambda_{\tau}$, and $p_{dp}(\cdot|x)$ and $p_{dn}(\cdot|x)$ denote the transition probabilities under the events that a beam from the mBS to the user is formed and is not formed, respectively, in the current slot, given the present state $x$. In each iteration, the above update equation \eqref{update_lambda} for $\lambda$ adjusts it so as to reduce the difference between the two arguments in the minimization function given in \eqref{Value_Function}. So from \eqref{update_lambda}, we observe that $\lambda$ converges to a value at which the controller obtains equal expected utilities by the formation and non-formation of a beam to the user when the state is $x$.

We find $V_{\lambda}(\cdot)$ using a linear system of equations, which are solved using the present value of $\lambda$ after each iteration of \eqref{update_lambda}. In particular, we solve the following system of equations for $V = V_{\lambda_{\tau}}$ and $\eta = \eta(\lambda_{\tau})$,  using the value of the parameter $\lambda = \lambda_{\tau}$:
\begin{align*}
   V(y) &= H(y) + P - \eta + \sum_{z}^{} p_{dp}(z|y) V(z), \, y > x, \\
   V(y) &= H(y) + \lambda - \eta + \sum_{z}^{} p_{dn}(z|y) V(z), \, y \leq x, \\
   V(0) &= 0.
\end{align*}
For a given state $x$, the value of $\lambda$ obtained after convergence of the iteration \eqref{update_lambda} is the Whittle index. To reduce the computational cost, we execute the iteration \eqref{update_lambda} only for a subset of the states $x$ and then, to find the Whittle indices for the remaining states, we use interpolation. Under the proposed Whittle index-based policy, the $B$ users that have the $B$ smallest  Whittle indices are assigned beams by the mBS in each time slot.

\section{Other Scheduling Policies}\label{other_policies}
In this section, we describe some other scheduling policies, whose performance is compared with that of the proposed Whittle index-based policy via simulations in Section \ref{simulations}. Recall that for $i \in \{1, 2, \ldots, K\}$, $X_n^i$ is the queue length of user $i$ in time slot $n$.

\subsection{Longest-Queue-First (LQF)}
     Under this policy, the $B$ users that have the $B$ longest queue lengths are selected in time slot $n$ and beams are assigned to them. Ties are broken at random.
    
\subsection{Max-Weight Scheduling (MWS)}
   Under this policy, the $B$ users that have the $B$ largest products, $X_n^i d_i$, of queue length, $X_n^i$, and probability of the channel quality being good, $d_i$, are selected in time slot $n$ and beams are assigned to them. Ties are broken at random.
    
\subsection{Weighted Fair Queuing (WFQ)}
   Under this policy, a positive weight $w_i$ is assigned to the queue of user $i \in \{1, \ldots, K\}$. Consider the probability distribution over the $K$ users in which the probability of user $i$ is $\frac{w_i}{\sum_{j=1}^{K} w_j}$. In each time slot, users are repeatedly selected  using this probability distribution until $B$ distinct users have been selected, and beams are formed with the selected $B$ users. We have assumed that $w_i = H_i(1)$ for $i \in \{1, 2, \ldots, K\}$, i.e., the weight of the queue of user $i$ equals its holding cost when there is $1$ packet in its queue.
    
\subsection{Random Scheduling}
    Under this policy, in each time slot $n$, $B$ users are selected uniformly at random out of the $K$ users, and beams are formed with the selected users.

\section{Simulations}\label{simulations}
In this section, we evaluate the performance of the proposed Whittle index-based policy and compare it with that of the LQF, MWS, WFQ, and random scheduling policies described in Section \ref{other_policies} through extensive MATLAB simulations. We evaluate the performance in terms of the following parameters: long-run expected average holding cost incurred at the mBS for the queues of all the users, average delay of packets, and the average number of active beams per time slot. The delay of a packet is the number of time slots required for the packet to be successfully transmitted after its arrival into a queue at the mBS. The average of the delays of all the packets of all the users in a simulation run is the average delay. Note that in a time slot, if one or more beams out of the $B$ available beams are not assigned to any user (this would typically happen when some of the queues are empty), then they can be switched off or deactivated, thereby saving energy. Hence, the average number of active beams per time slot is a measure of energy efficiency-- the lower the average number of active beams per time slot, the higher the energy efficiency and vice versa.   

We assume that the queue corresponding to each user at the mBS is initially empty. Each simulation runs for a period $(T)$ of $20,000$ time slots. Let $\textbf{d} = [d_1, d_2, \ldots, d_K]$, $\textbf{a} = [a_1, a_2, \ldots, a_K]$, and $\textbf{P} = [P_1, P_2, \ldots, P_K]$, where $d_i, a_i$ and $P_i$ are as defined in Section \ref{system_model}. We have also considered that the function $H_i(x)=q_i x^2$ for $i \in \{1,\ldots,K\}$, where $q_i$ is a constant. Let $\textbf{q} = [q_1, q_2, \ldots, q_K]$. The parameter `buffer size' equals the maximum number of packets that can be stored in a queue.

\begin{figure}[!t]
\centering
\subfloat[]{\includegraphics[width=0.24\textwidth]{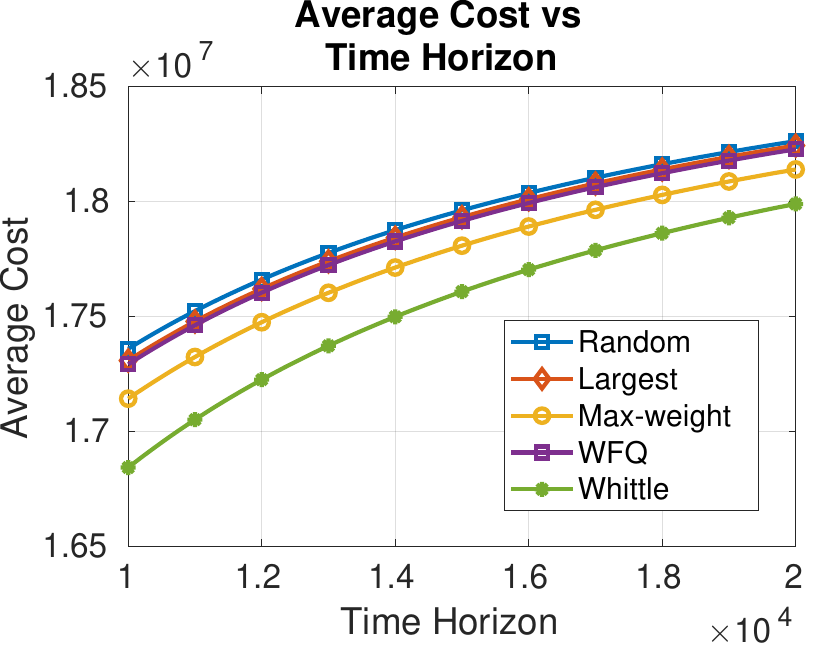}%
\label{fig:3a}}
\hfil
\subfloat[]{\includegraphics[width=0.24\textwidth]{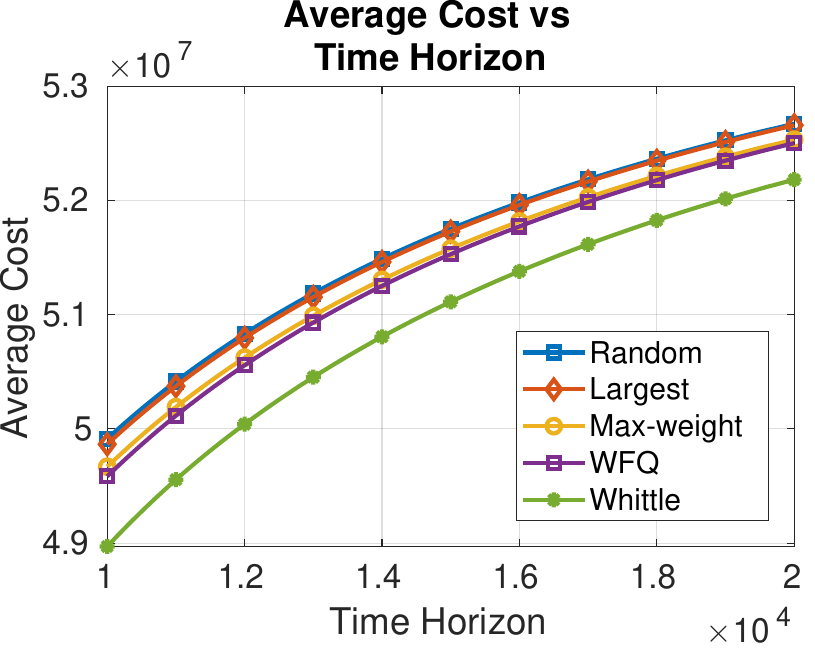}%
\label{fig:3b}}
\caption{The figures compare the average costs achieved under the five beam scheduling policies. The following parameter values are used for figure (a): $K=6$, $B=4$, buffer size $=400$, $\textbf{d}=[0.35,0.33,0.31,0.29,$ $0.27,0.25]$, $\textbf{a} = [0.55,0.52,0.49,0.46,0.43,0.4]$, $\textbf{P} = [60,55,50,45,40,$ $35]$, and $\textbf{q} = [30,26,22,18,14,10]$. The following parameter values are used for figure (b): $K=4$, $B=3$, buffer size $=500$, $\textbf{d}=[0.34,0.3,0.28,$ $0.32]$, $\textbf{a} = [0.58,0.56,0.57,0.55]$, $\textbf{P} = [87,74,62,49]$, and $\textbf{q} = [90,60,44,28]$.}
\label{fig:3}
\end{figure}

Fig. \ref{fig:3} plots the long-run expected average costs achieved under the five beam scheduling policies in time slots $10,001$ to $20,000$. The results are plotted for different values of the number of users, $K$, the number of beams, $B$, buffer size, $\textbf{d}$, $\textbf{a}$, $\textbf{P}$, and $\textbf{q}$. Fig. \ref{fig:3} shows that for all the parameter values considered, the proposed Whittle index-based policy achieves the lowest long-run expected average cost and outperforms all the other policies.

\begin{figure}[!t]
\centering
\subfloat[]{\includegraphics[width=0.241\textwidth]{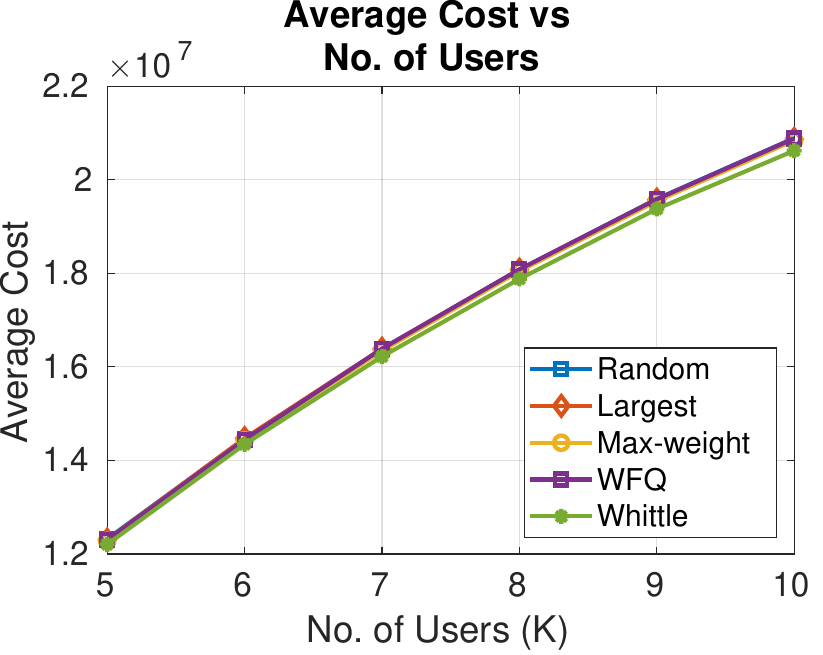}%
\label{fig:4a}}
\hfil
\subfloat[]{\includegraphics[width=0.241\textwidth]{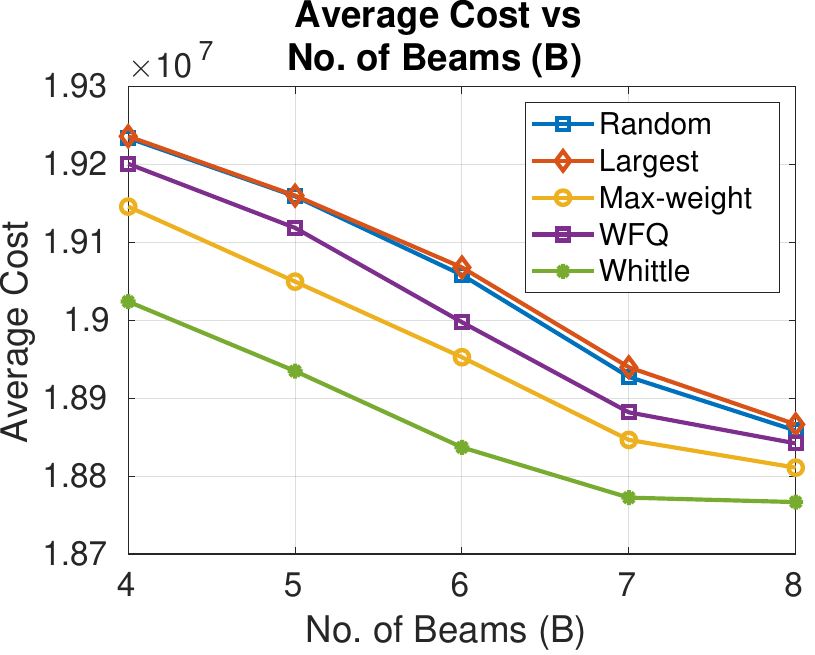}%
\label{fig:4b}}
\caption{The figures compare the average costs achieved under the five beam scheduling policies. Buffer size $=200$ for both the plots. The following parameter values are used for figure (a): $B=4$, and different values of $K$ varying from $5$ to $10$. For $K=5$, the following parameter values are used: $\textbf{d}=[0.30,0.28,0.29,0.31,0.28]$, $\textbf{a} = [0.52,0.51,0.5,$ $0.49,0.48]$, $\textbf{P} = [60,57,54,51,48]$, and $\textbf{q} = [80,75,70,65,60]$. For every subsequent addition of the $i^{th}$ user, $i \in \{6, \ldots, 10\}$, the values of $d_i$, $a_i$, $P_i$, and $q_i$ are selected as $0.28 \times (i \mod 2) + 0.29 \times ((i+1) \mod 2)$, $0.53-0.01i$, $63-3i$, and $85- 5i$, respectively, where $x \mod y$ denotes the remainder when $x$ is divided by $y$. The following parameter values are used for figure (b): $K=9$, $\textbf{d}=[0.25,0.241,0.231,0.222,0.213,0.204,$ $0.195,0.186,0.177]$, $\textbf{a} = [0.55,0.545,0.54,0.535,0.53,0.525,0.52,0.515,0.51]$, $\textbf{P} = [120,$ $110,100,90,80,70,60,50,40]$, $\textbf{q} = [90,82,74,66,58,50,42,34,26]$, and different values of $B$ varying from $4$ to $8$.}
\label{fig:4}
\end{figure}

Figs. \ref{fig:4a} and \ref{fig:4b} show the average costs versus the number of users ($K$) and the number of beams ($B$), respectively. In both of these plots, the proposed policy outperforms the other four beam scheduling policies. Intuitively, as $K$ increases, the mBS needs to serve a larger number of users with the same number of available beams. This explains why the average costs in Fig. \ref{fig:4a} increase as $K$ increases. Similarly, as $B$ increases, the mBS is able to simultaneously serve more users in each time slot. Hence, the average costs in Fig. \ref{fig:4b} decrease as $B$ increases. 

\begin{figure}[!t]
\centering
\subfloat[]{\includegraphics[width=0.241\textwidth]{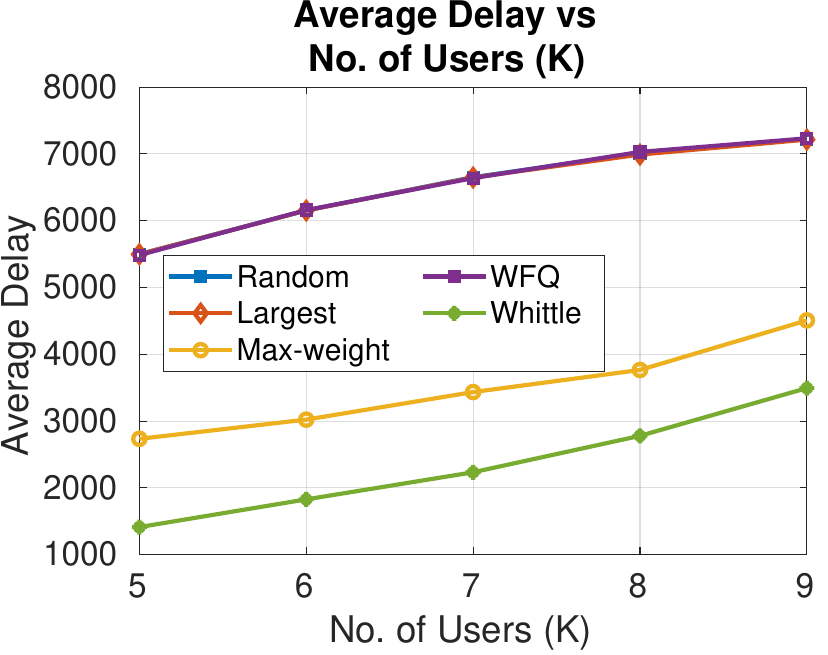}%
\label{fig:5a}}
\hfil
\subfloat[]{\includegraphics[width=0.241\textwidth]{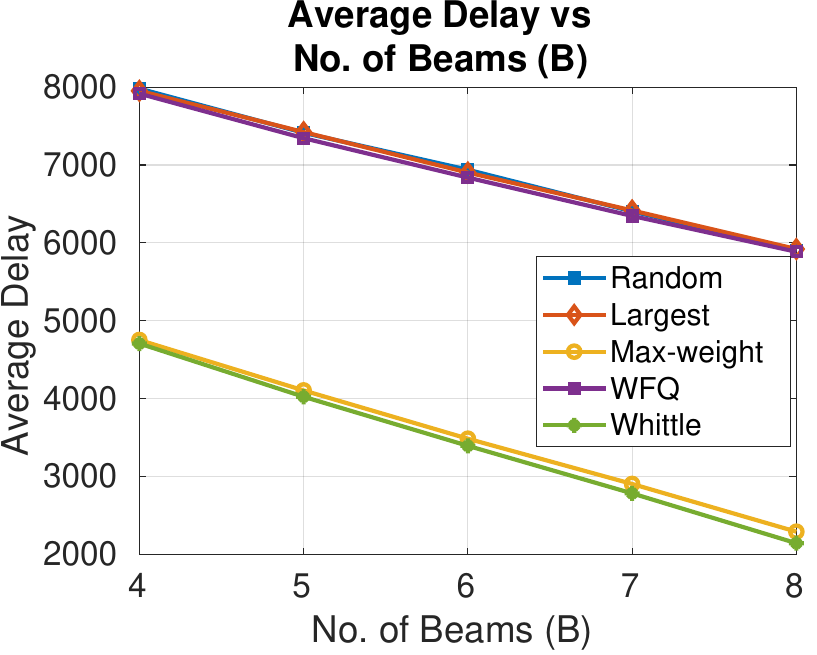}%
\label{fig:5b}}
\caption{The figures compare the average delays achieved under the five beam scheduling policies. Buffer size $=200$ for both the plots. The following parameter values are used for figure (a): $B=4$, and different values of $K$ varying from $5$ to $9$. For $K=5$, the following parameter values are used: $\textbf{d}=[0.29,0.285,0.28,0.275,0.27]$, $\textbf{a} = [0.56,0.53,$ $0.50,0.47,0.44]$, $\textbf{P} = [56,52,48,44,40]$, and $\textbf{q} = [82,78,74,70,66]$. For every subsequent user $i \in \{6, \ldots, 9\}$, the values of $d_i, a_i, P_i$, and $q_i$ are selected as $0.295-0.005i$, $0.59-0.03i$, $60-4i$, and $86-4i$, respectively. The following parameter values are used for figure (b): $K=9$, $\textbf{d}=[0.28, 0.272,0.253,0.243,0.231,0.222,0.21,0.196,0.187]$, $\textbf{a} = [0.505,0.504,0.503,0.502,0.503,0.502,0.503,0.502,0.503]$, $\textbf{P} = [60,55,50,45,40,35,30,25,20]$, $\textbf{q} = [85,77,69,61,53,45,37,$ $29,21]$, and different values of $B$ varying from $4$ to $8$.}
\label{fig:5}
\end{figure}

Figs. \ref{fig:5a} and \ref{fig:5b} show the average delays versus the number of users ($K$) and the number of beams ($B$), respectively. The reasons behind the increase in the average delays with $K$ in Fig. \ref{fig:5a} and their decrease with $B$ in Fig. \ref{fig:5b} are the same as those stated in the previous paragraph. Again, Fig. \ref{fig:5} shows that the proposed policy outperforms all the other policies.

Tables \ref{tab:active_beams_vs_B} and \ref{tab:active_beams_vs_K} show the average numbers of active beams per slot versus  $B$ and  $K$, respectively. From the tables, it is clear that the proposed Whittle index-based policy has the lowest average number of active beams, and hence the highest energy efficiency, among all the policies. Intuitively, this is because under the Whittle index-based policy, packets are served quickly after they arrive; hence, there are more empty queues per time slot on average, allowing the mBS to deactivate more beams per time slot on average than under the other policies. 

In summary, our simulations show that the proposed Whittle index-based policy outperforms the other four policies in terms of the average cost, average delay, as well as energy efficiency. 

\begin{table}
 \begin{center}
 \caption{The table shows the average  numbers of active beams per slot for the five beam scheduling policies versus $B$. The following parameter values are used: buffer size $=100$, $K=20$, different values of $B$ varying from $8$ to $16$, $\textbf{d}=[0.35,0.341,0.332,0.323,0.314,0.305,0.296,0.287,0.278,0.269,$ $0.260,0.251,0.242,0.233,0.224,0.215,0.206,0.197,0.188,0.179]$, $\textbf{a} =[0.65,0.645,0.64,0.635,0.63,0.625,0.62,0.615,0.61,0.605,0.60,$ $0.595,0.59,0.585,0.58,0.575,0.57,0.565,0.56,0.555]$, $\textbf{P} = [200,$ $190,180,170,160,150,140,130,120,110,100,90,80,70,60,50,40,$ $30,20,10]$, and $\textbf{q} = [174,166,158,150,142,134,126,118,110,$ $102,94,86,78,70,62,54,46,34,28,20]$.}
 \label{tab:active_beams_vs_B}
 \scalebox{0.9}{
 \begin{tabular}{|c|c|c|c|c|c|}
  \hline
  \multirow{2}{1.4cm}{\centering No. of Beams (B)} & \multicolumn{5}{c|}{Average No. of Active Beams} \\
  \cline{2-6}
  & Random & LQF & MWS & WFQ & Whittle \\
  \hline 
    8 & 7.9994 & 7.9996 & 7.9996 & 7.9995 & 7.9989\\ \hline
    9 & 8.9992 & 8.9995 & 8.9995 & 8.9993 & 8.9987\\ \hline
    10 & 9.9993 & 9.9995 & 9.9995 & 9.9994 & 9.9987\\ \hline
    11 & 10.9985 & 10.9994 & 10.9994 & 10.9986 & 10.9978\\ \hline
    12 & 11.9984 & 11.9994 & 11.9994 & 11.9984 & 11.9979\\ \hline
    13 & 12.9985 & 12.9993 & 12.9993 & 12.9988 & 12.9978\\ \hline
    14 & 13.9988 & 13.9992 & 13.9992 & 13.9990 & 13.9983\\ \hline
    15 & 14.9986 & 14.9991 & 14.9991 & 14.9985 & 14.9977\\ \hline
    16 & 15.9985 & 15.9990 & 15.9990 & 15.9976 & 15.9968\\ \hline
  \end{tabular}}
 \end{center}
\end{table}

\begin{table}
 \begin{center}
 \caption{The table shows the average  numbers of active beams per slot for the five beam scheduling policies versus $K$. The following parameter values are used: buffer size $=100$, $B=15$, and different values of $K$ varying from $16$ to $25$. For $K=16$, the following parameter values are used:  $\textbf{d}=[0.74,0.735,0.73,0.725,0.72,0.74,0.735,0.73,0.725,0.72,0.74,$ $0.735,0.73,0.725,0.72,0.74]$, $\textbf{a} = [0.64,0.635,0.63,0.625,0.62,0.64,$ $0.635,0.63,0.625,0.62,0.64,0.635,0.63,0.625,0.62,0.64]$, $\textbf{P} = [60,$ $55,50,45,40,60,55,50,45,40,60,55,50,45,40,60]$, and $\textbf{q} = [40,35,$ $30,25,20,40,35,30,25,20,40,35,30, 25,20,40]$. For every subsequent addition of the $i^{th}$ user, $i \in \{17, \ldots, 25\}$, the values of $d_i, a_i, P_i$, and $q_i$ are selected as $0.74-0.005 \times ( (i - 1) \mod 5)$, $0.64-0.005 \times ( (i - 1) \mod 5)$, $60-5 \times ( (i - 1) \mod 5)$, and $40-5 \times ( (i - 1) \mod 5)$, respectively.}
 \label{tab:active_beams_vs_K}
 \scalebox{0.9}{
 \begin{tabular}{|c|c|c|c|c|c|}
  \hline
  \multirow{2}{1.2cm}{\centering No. of Users (K)} & \multicolumn{5}{c|}{Average No. of Active Beams} \\
  \cline{2-6}
  & Random & LQF & MWS & WFQ & Whittle \\
  \hline 
    16 & 13.7877 & 13.7858 & 13.7846 & 13.7782 & 13.7670\\ \hline
    17 & 14.6887 & 14.6869 & 14.6834 & 14.6756 & 14.683\\ \hline
    18 & 15.4796 & 15.4777 & 15.4757 & 15.4664 & 15.4552\\ \hline
    19 & 16.1015 & 16.1001 & 16.0981 & 16.0881 & 16.0792\\ \hline
    20 & 16.9984 & 16.9976 & 16.9955 & 16.9864 & 16.9789\\ \hline
    21 & 17.7675 & 17.7661 & 17.7643 & 17.7557 & 17.7429\\ \hline
    22 & 18.3356 & 18.3334 & 18.3313 & 18.3249 & 18.3161\\ \hline
    23 & 18.6088 & 18.6063 & 18.6046 & 18.5944 & 18.5851\\ \hline
    24 & 19.1119 & 19.1110 & 19.1101 & 19.1089 & 19.0998\\ \hline
    25 & 19.4875 & 19.4865 & 19.4854 & 19.4788 & 19.4692\\ \hline
  \end{tabular}}
 \end{center}
\end{table}

\section{Conclusions And Future Work}\label{conclusion}
We studied the problem of beam scheduling for DL transmissions in a single-cell mmWave network. We proved that the problem is Whittle indexable and proposed a strategy to compute the Whittle index of each user. Using extensive simulations, we showed that our proposed Whittle index-based beam scheduling policy significantly outperforms several beam scheduling policies proposed in prior work in terms of the average cost, average delay, as well as energy efficiency. A future research direction is to extend the results of this paper to scenarios where multiple packets may arrive into and depart from the queue of a user at the mBS in each time slot.

\bibliographystyle{IEEEtran}

\end{document}